# Direct Photochemical Patterning of Lithium Niobate Thin Films for Scalable Nonlinear Optical Metasurfaces

Rana Faryad Ali* and Guillermo Aguilar

J. Mike Walker '66 Department of Mechanical Engineering, Texas A&M University, College Station, TX, 77843 USA

* Email: ranafaryad.ali@tamu.edu

**Abstract**

Lithium niobate is one of the most sought-after materials for nanophotonic devices, including frequency converters, modulators, and quantum light sources. Integration of lithium niobate into optical devices, however, is hampered by significant top-down fabrication challenges due to its exceptional chemical resistance. Scalable fabrication methods that preserve material quality while reducing fabrication complexity and cost are, therefore, crucial to advancing lithium niobate devices. We present a photochemical metal-organic decomposition technique for the scalable patterning of lithium niobate at ambient conditions, eliminating the need for harsh etching conditions and cleanroom protocols. The method utilizes a solution of a custom-prepared photosensitive organometallic precursor as a negative photoresist. The UV light exposure of the thin films of the precursor through a photomask, followed by rinsing with ethanol, yields amorphous patterns, which transform into crystalline lithium niobate after a calcination step. This method enables a scalable fabrication of a range of complex geometric shapes with a feature resolution down to 30 µm. The patterned lithium niobate structures exhibit a tunable second harmonic generation activity with an isotropic optical response. This approach offers a scalable and low-cost pathway for manufacturing lithium niobate photonics and the potential to fabricate other materials (e.g., barium titanite and lithium tantalate).

**Introduction**

Recent research in nanophotonics has focused on leveraging nonlinear optical processes [e.g., second harmonic generation (SHG)] to control and transform light into optical devices. These non-linear light-matter interaction processes find applications in optical modulators, waveguides, transducers, color filters, photodetectors, photon sources, and ultrafast optical switches.[1–3] Due to the inherent weakness of the SHG, however, contemporary applications of SH-active materials predominantly rely on high-intensity optical sources and bulky crystalline materials.[4–6] Emerging strategies (e.g., advanced material systems and fabrication designs) are, therefore, being developed to enhance the intrinsically weaker SHG process.[7] Nanoparticles of the nonlinear optical materials offer a number of strategies to enhance SHG, however, they have limited use due to challenges in their integration into optical devices.[8,9] The recent development of meta-optics represents a promising fabrication design direction for enhancing weaker SHG processes and miniaturizing nanophotonics.[10] Metastructures consist of periodic arrangements of nanostructures (i.e., meta-atoms). These meta-atoms can modulate the phase of light (e.g., as in metalenses), and/or enhance light-matter interactions (e.g., as in resonant metasurfaces).[11] This controlled and tunable arrangement of meta-atoms enables materials with new capabilities, including, but not limited to, enhanced SHG response.[10,11] While this field holds a great promise, it is still in its early stages for nonlinear optical materials due to the challenges of fabrication and integration of material systems.

The choice of materials is also critical for the design and applications of nonlinear optical metasurfaces. A number of material systems, such as silicon (Si), gallium arsenide (GaAs), and gallium phosphide (GaP), are commonly used to design nonlinear metasurfaces due to their ease of fabrication and larger second-order nonlinear susceptibility.[12] These material systems, however, suffer from high losses in the visible part of the spectrum and begin to absorb light in the near-infrared (NIR) range. These limitations restrict the utility of these material systems in developing low-loss nonlinear optical devices operating in the visible region of the spectrum.[12,13] Meta-optics components, therefore, require highly transparent materials, have a relatively large second-order nonlinear susceptibility, and possess a high refractive index.

Non-centrosymmetric metal-oxides [e.g., lithium niobate (LN) and barium titanite (BaTiO$_3$)] present a promising alternative platform for meta-optics. These materials offer substantial second-order susceptibility coefficients, exceptional optical damage thresholds, and broad transparency windows.[14,15] Among these platforms, lithium niobate (LN or LiNbO$_3$) turns out to be particularly suitable for nonlinear meta-optics and devices. It has been described as "the silicon of photonics because it offers a rare combination of advantageous properties.[7,13,16,17] These properties include, but are not limited to, the largest second-order susceptibility tensor ($\chi^{(2)}$), broad transparency (400 nm to 5 μm), long-term chemical stability, and higher optical damage threshold.[18] This makes LN one of the key photonic materials with the potential to expand access to ultrawide parts of the nonlinear optical spectrum and support the next generation of scientific breakthroughs. Despite these exceptional material properties, the practical implementation of LN in advanced photonic devices has been historically constrained by significant fabrication challenges.

The materials properties (e.g., chemical and physical inertness) of LN, while beneficial for device stability and longevity, create substantial difficulties during top-down fabrication processes.[7,12,13] Traditional top-down fabrication approaches (e.g., e-beam lithography, photolithography, and ion beam milling) require multi-step deposition/development, harsh etching processes, and specialized equipment.[19,20] This complexity of methods limits the utility of top-down methods for the scalable and high-throughput fabrication of LN meta-optics and devices. Recently, bottom-up approaches (e.g., solution-phase chemistries) have been developed to pattern and fabricate LN meta-optics and devices.[21,22] For example, soft nanoimprinting lithography (SNIL) is one such approach that offers a potential solution for the scalable patterning of LN meta-optics.[22] The SNIL, however, has several limitations that limit its practical utility. These limitations of SNIL include a lack of ability to pattern complex shapes and larger structures (e.g., waveguides), and is prone to reproducibility issues when processing conditions are not carefully controlled (e.g., mold alignment).[23,24] These limitations mean that SNIL, while promising, is not yet a general solution for the scalable and high-throughput fabrication of LN and other nonlinear optical metasurfaces.

In this paper, we introduce a photochemical metal-organic decomposition (PMOD) method for the scalable and flexible fabrication of customized LN patterns for nonlinear optical metasurfaces. This PMOD method enables the fabrication of intricate patterns of LN at ambient conditions and eliminates the need for clean rooms, harsh etching, and/or lift-off processes. We prepare a photosensitive lithium-niobium organometallic precursor, which acts as a negative photoresist to pattern LN structures. An exposure of the spin-cast thin films of the precursor to ultraviolet (UV) light through a photomask results in their photodecomposition in the exposed regions. The development of UV irradiated thin films with ethanol rinses away the soluble precursor that remained in the regions blocked from UV exposure and leads to the formation of well-defined patterns of LN. This PMOD method offers a high degree of versatility and enables the scalable fabrication of a range of patterns (e.g., logos, lines, squares, circles, and complex shapes) with minimum feature sizes as low as 30 µm. Detailed investigation of the patterned materials shows the formation of a high-quality, phase-pure, porous, and polycrystalline LN thin-film patterns sought after nonlinear optical metasurfaces. A comprehensive analysis of the SHG response indicates the formation of SH-active LN patterns. The patterned thin films of LN exhibit polarization-independent SHG, a characteristic of randomly oriented polycrystalline media. This work addresses the current challenges in the scalable fabrication of LN and can be extended to other material systems (e.g., $BaTiO_3$, $LiTaO_3$), showing considerable potential to enable low-cost and scalable fabrication of optical devices.

**Results and Discussion**

This paper describes a scalable fabrication technique to pattern lithium niobate (LN) structures with tunable sizes and shapes for nonlinear optics. We developed a photochemical metal-organic decomposition (PMOD) method. Unlike conventional fabrication methods (e.g., e-beam lithography and ion milling), this PMOD method enables the patterning of LN structures without the need for cleanrooms, harsh etching conditions, and/or multiple processing steps.[25,26] Our developed PMOD method to fabricate customized patterns of LN include five steps (**Figure 1**): (i) preparation of a photosensitive precursor; (ii) coating of the photosensitive precursor onto substrates; (iii) UV exposure at selected regions via a predesigned photomask; (iv) developing with a non-aqueous solvent (i.e., ethanol) to remove unexposed precursor; and (v) calcination of amorphous patterns to obtain crystalline materials.

The PMOD approach uses a solution-phase chemistry in which a photosensitive precursor solution (e.g., aqueous or non-aqueous) is a key component of the process. The choice of precursors, photosensitive ligands, and solvents dictates the efficacy of the presented patterning method. We prepared a non-aqueous photosensitive organometallic precursor containing lithium and niobium species in ethanol (**Figure 1a**). The design of non-aqueous photosensitive precursors was motivated by the goal of achieving better control over reaction chemistry and the quality of LN patterns.[27] For this purpose, we selected lithium niobium ethoxide [LiNb(OEt)$_6$], which offers a precise 1:1 stoichiometry between lithium and niobium and offers an ease to readily replace ethoxide ligands with photosensitive ligands. Many oxide materials, including LN, scatter and/or absorb in the UV region of light. We decided, therefore, to utilize a photosensitive ligand that yields a precursor sensitive to longer UV wavelengths (e.g., > 300 nm).[28] Among the many ligand systems available to yield photosensitive precursors, we used benzoylacetone (i.e., a β-diketone) as a ligand due to several key advantages over other alternatives.[28–31] For example, compared to other ligands, such as acetylacetone, BzAc exhibits superior photosensitivity due to its conjugated π-electron system in the phenyl ring, which facilitates π → π* and n → π* transitions, enabling efficient photolysis of the complex. Additionally, BzAc provides an optimal balance between photosensitivity, chemical stability (e.g., steric hindrance), commercial availability, and cost-effectiveness, even compared to ligands that absorb longer-wavelength UV light.[28–31]

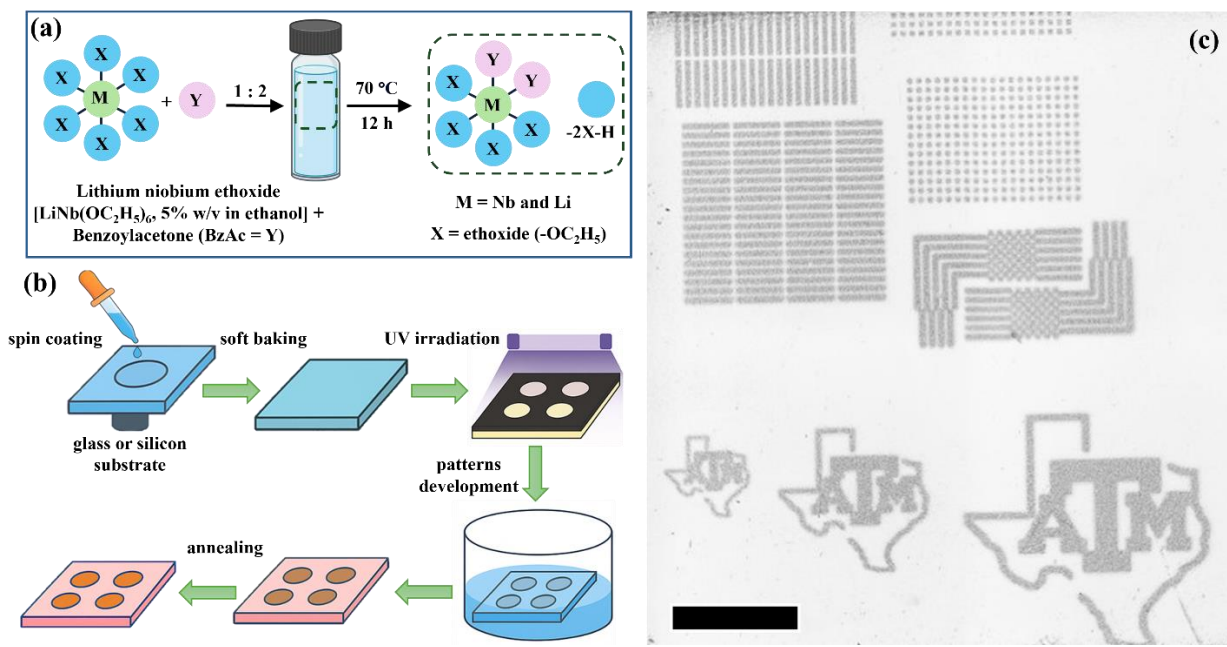

**Figure 1.** Direct photochemical metal-organic deposition (PMOD) lithography for scalable patterning of lithium niobate (LN) thin films. (a) Schematic illustration of the preparation of the non-aqueous photosensitive precursor solution, which acts as a negative photoresist for patterning LN. (b) Process flow of the PMOD lithography used to pattern LN on silicon and/or glass substrates from the prepared precursor solution. (c) Optical microscopy image (scale bar = 5 mm) showing an array of crystalline LN patterns with varying designs and complexities fabricated simultaneously using PMOD lithography.

We first optimized the mole ratio of $LiNb(OC_2H_5)_6$ to BzAc. This optimization is critical to control the photosensitivity of the precursors, the material quality, the required UV exposure time, and the solubility of the films after exposure. The initial tests were focused on a 1:1 mole ratio between metal alkoxide and BzAc. The formation of a photosensitive complex between $LiNb(OC_2H_5)_6$ and BzAc proceeds through a ligand-exchange mechanism. Benzoylacetone ($C_6H_5$-CO-$CH_2$-CO-$CH_3$) exists predominantly in its enol tautomeric form ($C_6H_5$-CO-CH=C(OH)-$CH_3$) in solution. In this form, the enolic proton is readily deprotonated to form the BzAc anion (BzAc$^-$). This anionic species acts as a chelating ligand, coordinating to the metal centers to form thermodynamically stable six-membered chelate rings. [28–31] The proposed reaction mechanism can be represented as:

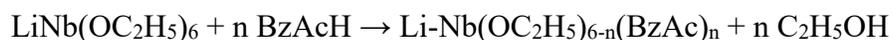

$$\text{LiNb(OC}_2\text{H}_5)_6 + \text{n BzAcH} \rightarrow \text{Li-Nb(OC}_2\text{H}_5)_{6-n}(\text{BzAc})_n + \text{n C}_2\text{H}_5\text{OH}$$

Here, n represents the number of ethoxide ligands replaced by BzAc⁻ ligands.

We acquired the UV-visible absorption spectra of the pristine starting materials and the mixed reaction solutions to investigate the formation of the precursor and optimize the mole ratio between metal alkoxide and BzAc (**Figure S1**). The pristine LiNb(OC$_2$H$_5$)$_6$ precursor exhibited a strong absorption band in the UV region at ~250 nm, which is characteristic of metal-to-ligand charge-transfer transitions. After the reaction with one mole of BzAc, however, the UV absorption spectrum showed a significant redshift, and a new absorption band appeared at ~320 nm. No sharp absorption edge was observed at ~365 nm for this reaction mixture, indicating the absence of n → π* electronic transitions. These transitions are critical for monitoring successful ligand-exchange reactions and the formation of metal-β-diketonate complexes.[28–31] This spectral behavior suggested incomplete chelation or weak electronic coupling between the metal centers and the benzoylacetonate (BzAc⁻) ligands. When the molar ratio between the metal alkoxide and the ligand was increased to 1:2, however, a distinct absorption edge appeared in the UV-visible spectrum at ~365 nm. This new feature indicates the presence of n → π* electronic transitions, confirming a successful ligand-exchange reaction and the incorporation of BzAc into the metal alkoxide framework.[28–31] This spectral evolution demonstrates that a 1:2 stoichiometric ratio is optimal for achieving complete chelation and imparting the desired photosensitive characteristics. The enhanced absorption intensity and the appearance of the characteristic absorption edge at 365 nm further suggest that sufficient BzAc⁻ ligands are coordinated to the metal centers, forming an extended conjugated system essential for efficient photochemical activity. [28–31] The resulting precursor undergoes photodecomposition initiated by the absorption of 365 nm UV photons, whose energy corresponds to the π → π* and n → π* electronic transitions of the metal complex.

Upon photoexcitation, the complex is promoted to an electronically excited state, causing redistribution of electron density within the chelate ring.[28,30,31] In this state, electron density shifts toward the oxygen atoms of the benzoylacetonato ligand, weakening the coordination bonds between the metal centers and the ligand oxygen atoms. The weakened metal–ligand bonds subsequently dissociate, releasing BzAc or its photolysis products (e.g., reactive radical species and organic fragments). These reactive intermediates undergo intermolecular coupling reactions, forming crosslinked metal-oxygen-metal (M-O-M) networks that ultimately result in the development of an amorphous metal–oxygen framework.[28–31]

We used the prepared photosensitive lithium-niobium complexes and spin-cast them onto polished silicon wafers and glass substrates to test their ability to pattern LN (**Figure 1b**). After coating the precursors, the substrates were baked at 70 °C for 90 s to remove any residual solvent from the films. These deposited films containing the organo-lithium-niobium complex were exposed to 365 nm UV light through a photomask at ambient conditions. This process allowed us to investigate the photodecomposition in the regions exposed to UV light. After UV exposure, patterned films of amorphous LN were obtained by the selective removal of the unexposed regions through rinsing the substrates with ethanol. We used an exposure time of 80 min in our initial studies to test both 1:1 and 1:2 precursors for their ability to yield patterns. No thin-film or pattern formation was observed for the 1:1 ratio precursor in either the exposed or unexposed regions after development with ethanol. In contrast, distinct thin-film patterns were formed at the exposed regions for the precursor with a 1:2 ratio of metal alkoxide to BzAc. This analysis further showed that the precursor with a 1:2 ratio of alkoxide and BzAc can undergo photodecomposition to yield thin-film patterns. We further investigated the optimal UV exposure time for patterning LN thin films. An exposure time shorter than 50 min resulted in partially developed patterns, while an exposure time longer than 90 min produced overexposed features. These observations indicated that an exposure time of ~50 min was optimal for the photodecomposition of the precursor with a 1:2 ratio between metal alkoxides and BzAc. Finally, the amorphous patterned thin films were calcined at 650 °C for 45 min to yield crystalline LN samples.[7] In all subsequent studies, we used these optimized conditions to fabricate LN thin-film patterns.

For the initial proof-of-concept studies, we focused on relatively larger patterns to establish the fundamental capabilities of the PMOD technique under ambient conditions (**Figure 1c**). We designed test patterns containing lines with a width of 180 µm, squares with a side length of 180 µm, and circles with a diameter of 180 µm. These initial experiments successfully demonstrated the ability to pattern LN films across large substrates (e.g., > 3 inches) under ambient atmospheric conditions without requiring controlled environments and harsh etching conditions. Optical microscopy analysis revealed excellent pattern transfer fidelity. The patterns showed sharp edge definition and uniform film thickness across all patterned features. We further demonstrated the versatility of the approach by successfully fabricating complex patterns such as logos and interconnected geometric shapes (**Figure 1c**). These proof-of-concept patterns validated the capability of the method for scalable patterning of both simple geometric features and intricate designs. The method also accommodated multiple length scales and varying feature densities.

We characterized the crystal structure and phase of the patterned LN films by powder X-ray diffraction (XRD) techniques. We acquired the XRD patterns of LN films before and after calcination (**Figure 2a and Figure S2**). The XRD patterns of the calcined samples were indexed and correlated well with the reported trigonal LN reference (space group *R3c*, JCPDS no. 020-0631).[7,32] The powder XRD patterns of non-calcined films, however, confirmed the presence of an amorphous material. These analyses indicated the formation of a pure, rhombohedral, crystalline LN product after the calcination step. We further characterized the composition, purity, and phase of the patterned LN films by Raman spectroscopy (**Figure 2b and Figure S3**). Raman spectra were acquired for both calcined and non-calcined films and compared to evaluate the phase of the materials. The Raman spectra of the calcined LN films were consistent with those reported for LN materials. This indicated the formation of a crystalline rhombohedral product. For non-calcined samples, however, a broad Raman band was observed, indicating the presence of amorphous material. The Raman spectra of the calcined LN films were consistent with those reported for LN materials, indicating the formation of a crystalline rhombohedral phase.[7,32] The Raman bands below 250 cm$^{-1}$ (E-TO) were assigned to the deformation of the NbO$_6$ framework, and the band centered at ~600 cm$^{-1}$ (A1-TO) corresponded to the symmetric stretching of Nb-O-Nb bonds in the rhombohedral LN crystals. Raman bands at ~430 cm$^{-1}$ and ~375 cm$^{-1}$ (E-TO) were associated with the bending modes of the Nb–O–Nb bond, while the Raman band above 800 cm$^{-1}$ was assigned to the antisymmetric stretching of the Nb–O–Nb bonds in the NbO$_6$ octahedra of LN. Characteristic Raman bands for LiNb$_3$O$_8$ (59, 79, 96, 542, 701, and 738 cm$^{-1}$) and Li$_3$NbO$_4$ (470, 520, 748, and 824 cm$^{-1}$) were absent in the Raman spectra of the calcined LN films. These Raman spectroscopy results indicated the formation of pure rhombohedral LN with proper stoichiometric composition.

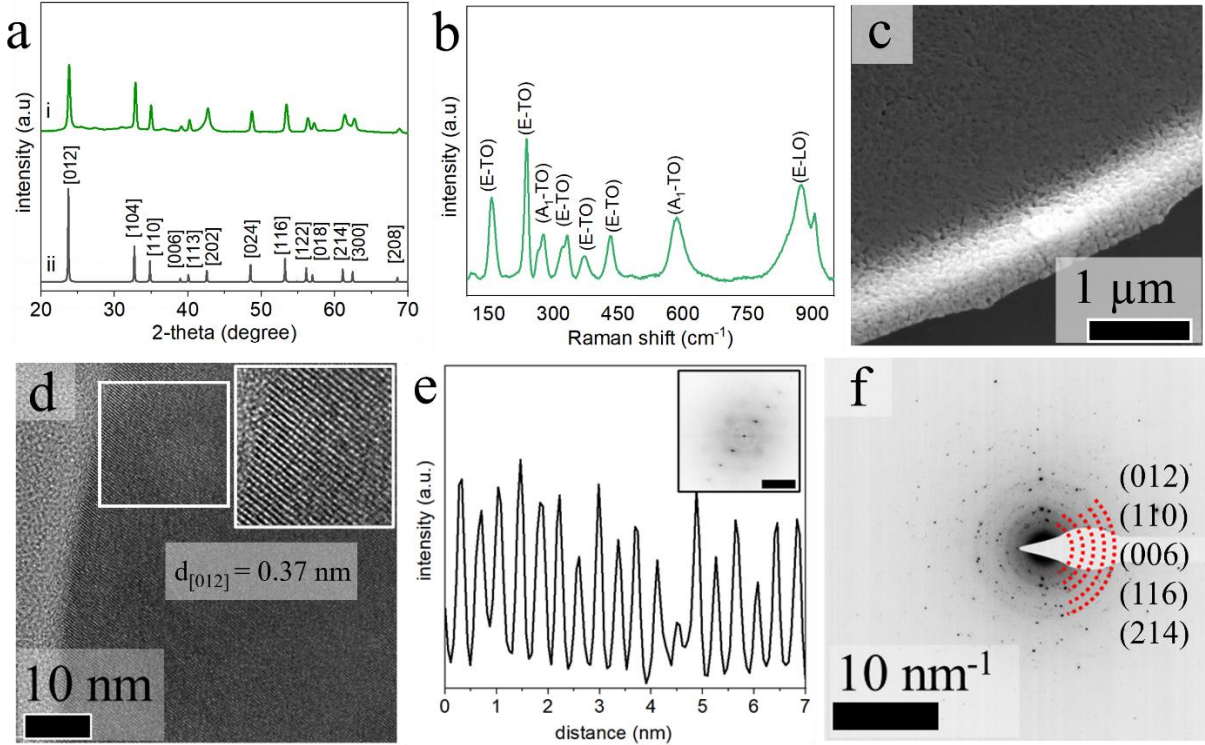

**Figure 2.** Structural and compositional characterization of photochemically patterned LN thin films. (a) Powder X-ray diffraction (XRD) patterns of calcined patterned LN thin films (i, green) and reference (ii, black; JCPDS No. 00-020-0631), confirming the rhombohedral crystal structure. (b) Raman spectroscopy analysis showing characteristic vibrational modes of phase-pure crystalline LN with no evidence of secondary phases. (c) Cross-sectional SEM image revealing porosity, polycrystallinity, and a film thickness of ~300 nm after calcination. (d) High-resolution transmission electron microscopy (HRTEM) image with a magnified inset showing lattice fringes with a d-spacing of 0.37 nm corresponding to the (012) planes of LN. (e) HRTEM lattice spacing profile and corresponding Fast Fourier Transform (FFT) analysis (scale bar = 5 nm⁻¹) further confirm the crystalline nature of LN. (f) Selected area electron diffraction (SAED) of LN grains displaying diffraction rings indexed to rhombohedral LN, further confirming its crystallinity and phase purity.

We measured the thickness of the calcined thin films of LN using the cross-sectional scanning electron microscopy (SEM) technique (**Figure 2c**). The thickness of the patterned films was ~300 nm after calcination at 650°C for 45 minutes. The SEM images further revealed that the films were porous in nature, with a network of interconnected and randomly oriented grains of LN. This random orientation and porosity can be attributed to the removal of organic components during the photochemical conversion and subsequent calcination process. To evaluate the crystalline structure and phase purity of the grains of the calcined LN films, we acquired transmission electron microscopy (TEM) analyses. We carefully peeled the patterned LN films to prepare samples for transmission electron microscopy. The high-resolution electron microscopy (HRTEM) analysis provided direct visualization of the atomic-scale structure and crystalline quality of individual grains within the calcined LN films (**Figure 2d and Figure 2e**). The measured lattice spacings from HRTEM images were 0.37 nm, corresponding to the (012) crystallographic planes of rhombohedral LN.[33] We further acquired the selected area electron diffraction (SAED) analysis to gain insights into the crystalline structure of the grains of the calcined films (**Figure 2f**). The SAED ring patterns were systematically indexed to identify the crystallographic planes and confirm the rhombohedral lithium niobate structure. The observed SAED patterns corresponded precisely to the expected values for rhombohedral LN (space group *R3c*), with the most prominent rings indexed to the (012), (110), (006), (116), and (214) reflections.[33] Compositional analysis by energy-dispersive X-ray spectroscopy (EDS) confirmed the presence of niobium (Nb) and oxygen (O) in the particles, and the lack of impurities of other elements (**Figure S4**).

We further systematically evaluated the resolution limits of the PMOD technique by a progressive reduction of feature sizes while maintaining consistent processing conditions. Test patterns of varying geometries, including lines, squares, and complex interconnected structures, were designed on a chrome mask. The width of the lines and squares was 140 μm, 110 μm, 90 μm, 70 μm, 60 μm, and 30 μm. This design enabled assessment of resolution dependence on feature geometry in the PMOD process. We simultaneously patterned all features on a single large substrate to demonstrate scalability and to ensure consistency in precursor composition, film coating, exposure dose, ambient conditions, and development procedure. Optical microscopy images of the patterns revealed distinct resolution-dependent behavior (**Figure 3**). We observed well-defined features as small as 70 μm for both line and square patterns. This demonstrates that the optimized parameters, including precursor composition, coating, exposure time, and development procedure, are well-tuned for fabricating patterns of various sizes. The interconnected structures also preserved their designed connectivity between lines and squares, indicating that the PMOD method can reliably reproduce complex geometries at these dimensions. The uniform development across different feature types further confirms that the exposure and development conditions were appropriately optimized for this size range.

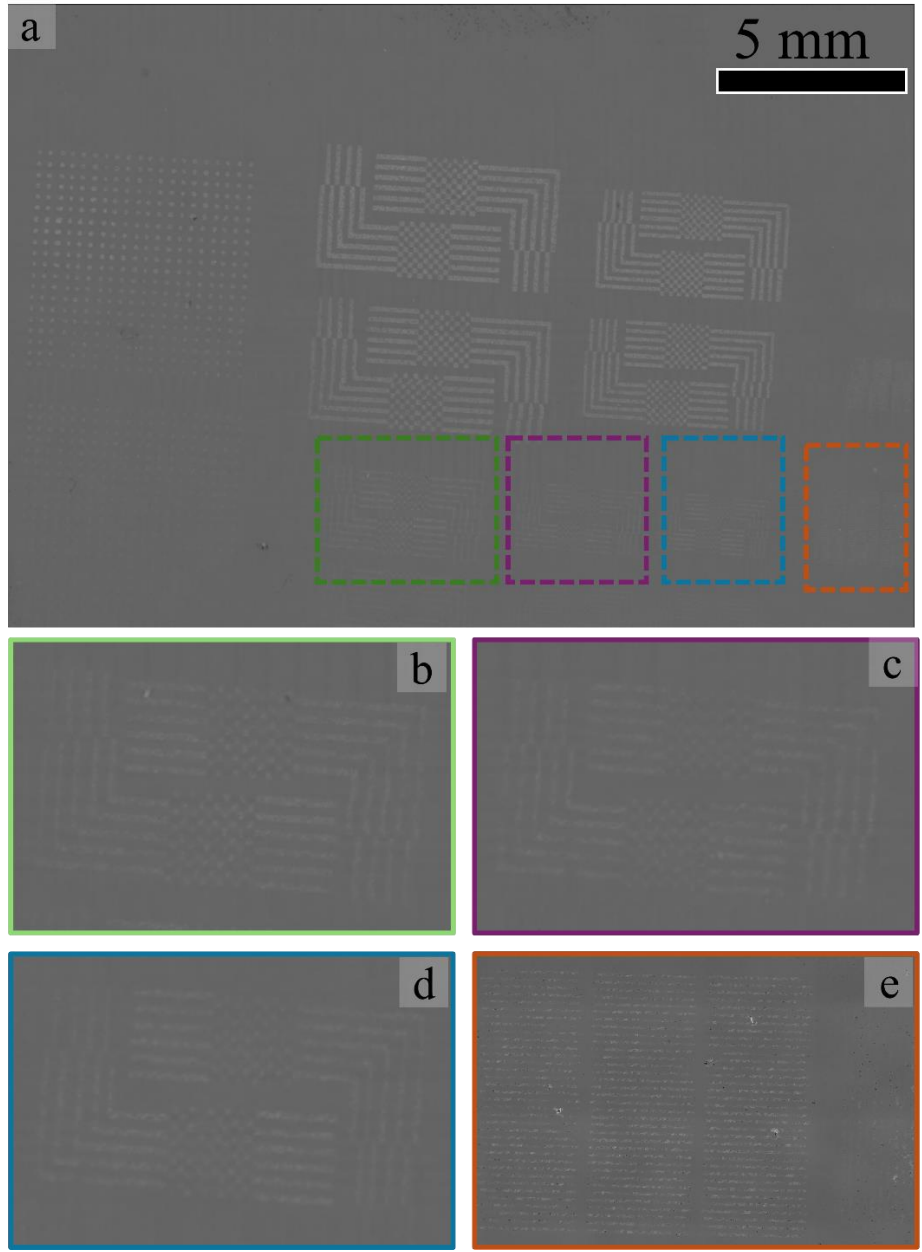

**Figure 3.** Optical microscopy analysis of photochemically patterned LN features showing resolution optimization. (a) Large-area stitched optical microscopy image demonstrating scalable patterning capability with various feature sizes and geometries, including lines, squares, circles, and complex interconnected shapes. Both complex interconnected shapes have a feature size of 140 and 110 µm. (b-e) Magnified views of optimized pattern regions: (b) 90 µm proof-of-concept features with excellent edge definition, (c) 70 µm patterns maintaining high fidelity, (d) 60 µm patterns demonstrating geometry-dependent resolution limits where lines develop successfully while squares have poor development, and (e) 30 µm lines demonstrating ultra-high resolution capability with minor non-uniformity at the resolution limit of this PMOD process.

At the 60 μm feature size, however, we observed a critical threshold behavior where pattern development became geometry-dependent (**Figure 3d**). While line features continued to develop successfully, maintaining their designed width and showing good edge sharpness, square features failed to develop properly. This geometry-dependent resolution limit can be attributed to several factors, such as the kinetics of development and/or variation in the dose of UV light.[34,35] For example, the corner regions of square features may receive reduced UV intensity due to diffraction at the photomask edges, leading to incomplete photochemical conversion of the precursor and subsequent development failure. We further observed that the PMOD method can pattern line features as narrow as 30 μm (**Figure 3e**). At this ultra-high-resolution limit, however, some line patterns exhibited artifacts, nonuniform material distribution, and increased edge roughness and discontinuities relative to their designed dimensions. This non-uniformity at 30 μm line width likely arises from the limited UV power and enhanced diffraction effects at smaller scales.[25,27,34,35] In addition, smaller line features possess a higher surface area-to-volume and perimeter-to-area ratio, which enhances solvent penetration during development and can lead to overdevelopment or pattern collapse.[34,35] We believe that the PMOD technique can be further improved for sub-30 μm patterning by optimizing key parameters such as UV exposure dose, lamp energy, development time, solvent composition, and precursor formulation.

We examined the optical second harmonic response of the LN patterns to evaluate their nonlinear optical (NLO) properties. The SHG emission from the patterns was analyzed using a commercial multiphoton microscope (**Figure 4a**). Nonlinear optical materials are widely used in various applications, including quantum light sources, frequency converters, ultrafast optical switches, and optical memory devices.[2,7,12,17,22] Second harmonic generation (SHG) is a second-order NLO process in which two incident photons of the same frequency (ω) combine to generate a single photon with twice the frequency (2ω) (**Figure 4b**). The SHG signal arises when light of sufficiently high intensity interacts with a material that lacks inversion symmetry. The oscillating electric field of the incident photons at frequency (ω) induces a nonlinear polarization in the medium, which upon relaxation emits a photon at twice the incident frequency (2ω).[2,7,12,17,22]

Recently, polycrystalline and porous structures of materials have attracted significant interest for their potential to address current challenges in nonlinear metasurfaces and metalenses.[7,7,22,32] The advantage of porous and polycrystalline materials lies in their ability to generate a SHG response without requiring phase-matching conditions, while maintaining a uniform signal output. In bulk NLO materials, where the crystal dimensions exceed the wavelength of the incident light, the SHG intensity becomes appreciable only under phase-matching conditions. Phase matching occurs when the generated SHG wave propagates through the material with the same phase velocity and in the same direction as the fundamental wave, resulting in constructive interference. In contrast, porous and polycrystalline NLO structures do not rely on strict phase matching to produce a detectable SHG signal. This behavior arises from the relatively small grain sizes compared to the wavelength of the incident photons.[7,7,22,32] As a result, the SHG generated within individual grains remains approximately in phase, allowing efficient scattering and emission along multiple directions.

We acquired input wavelength-dependent and power-dependent NLO spectra for the patterned LN thin films to ensure the patterns are SHG active and to verify the second-order nature of the detected signals. Power-dependent SHG signals were acquired by varying the power of the laser at the FW of 1,000 nm while imaging the areas of patterns on the substrate. False-colored images corresponding to the intensity and wavelength of the SHG response were obtained from the patterns (**Figure 5a**). The image emission matrix was averaged to yield a mean value in regions of interest selected to exclude the majority of the dark pixels. The selected regions of interest were maintained unchanged throughout the series of acquired images. We then plotted the mean values for the SHG signals against the laser power (**Figure 5b**). The intensity of the SHG signals increased with an increase in the laser power, reflecting the dependence of SHG on the pumping intensity. A second-order polynomial fit of the data points confirmed that the observed optical response is due to a second-order NLO process (i.e., SHG response). The patterned LN structures, therefore, are SH-active. We further demonstrated that the SHG response for patterned LN structures is tunable for a series of discrete fundamental wavelengths. (**Figure 5c**). The measured data points were normalized to their maximum intensity and fitted using a Gaussian function (i.e., spectroscopy function) in Origin Pro to estimate the peak position of each of the measured SHG responses. We observed a tunable SHG response at 420, 440, 460, and 480 nm when the patterns were excited with FWs of 840, 880, 920, and 960 nm, respectively. These results correlated well with the anticipated frequency doubling of the second-order NLO process of SHG, confirming the ability of the LN patterns to yield a tunable SHG response. To understand the role of polycrystallinity and randomly oriented grains of LN thin films, we also measured polarization-resolved SHG emission (**Figure 5d**). The near-circular polar plots of SHG intensity indicate an isotropic and uniform response across different polarization angles of the fundamental wavelength. This polarization-independent SHG behavior can be attributed to the polycrystalline nature and random grain orientation of the LN thin films, which eliminate polarization selection rules.[22,36] The polycrystalline LN patterns with randomly oriented grains, therefore, offer a key advantage by relaxing the stringent phase-matching conditions of single crystals, enabling broadband SHG responses independent of the pump polarization. The demonstrated scalable fabrication of SH-active LN patterns could have potential utility in a wide range of nonlinear photonic devices (e.g., waveguides, frequency converters, electro-optic modulators, and quantum light sources).

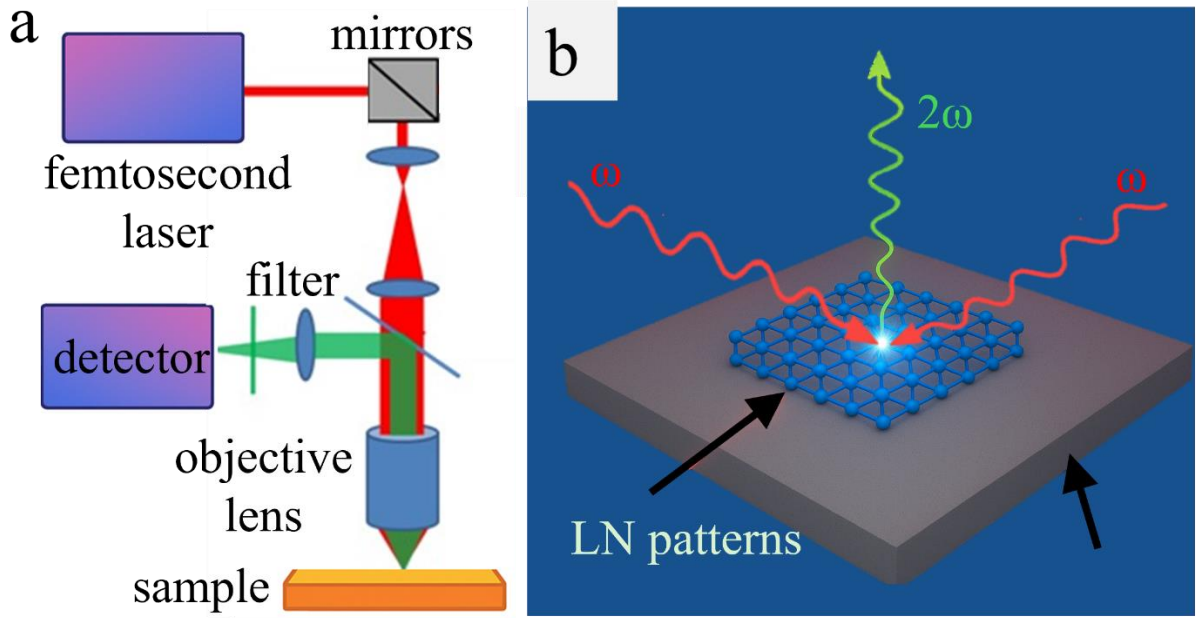

**Figure 4.** (a) Schematic of the microscope configuration used to characterize the second harmonic generation (SHG) response of patterns. This microscope was operated in a reflection mode and was equipped with a femtosecond laser, a half-wave plate, and objective lenses of different magnifications. (b) Illustration of the second-order nonlinear light-matter interactions responsible for SHG, in which two photons of the same wavelength interact with the material to produce a single photon of twice the frequency (or half the wavelength).

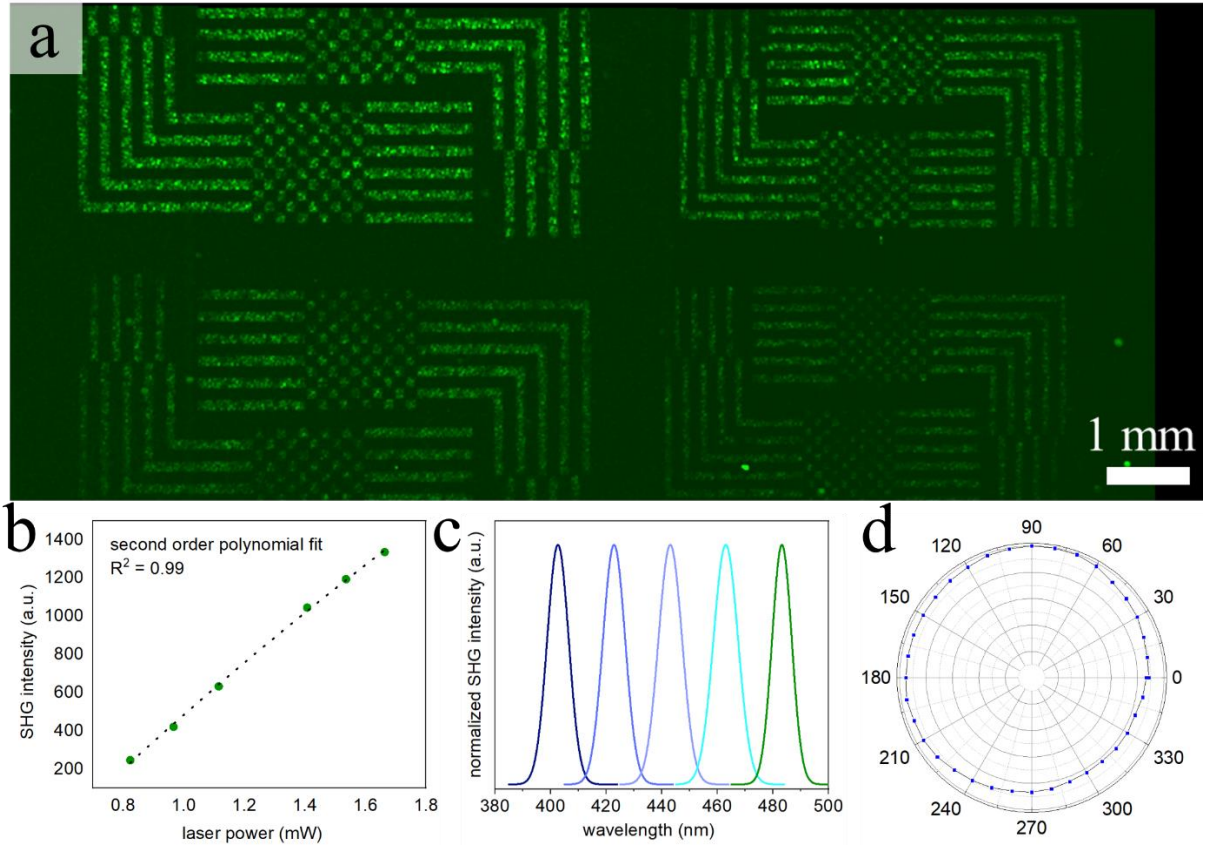

**Figure 5.** Second harmonic generation (SHG) characterization of photochemically patterned LN films. (a) Pseudo-color SHG microscopy image of patterned LN features showing the SHG response across the patterned regions. (b) Power dependence of SHG intensity demonstrating a second-order polynomial fit of the data points, confirming a second-order nonlinear process. (c) Wavelength tunability of the SHG response across fundamental wavelengths from 840-960 nm, demonstrating broad spectral tunability. (d) Polarization dependence plot showing an isotropic SHG response across all polarization angles, confirming random crystallographic orientation and polarization-independent behavior.

## Conclusion

In this work, we introduced a PMOD method for the scalable and direct patterning of SH-active LN thin films. This bottom-up method operated under ambient conditions and eliminated the need for cleanroom facilities and/or harsh etching processes to fabricate patterns of chemically inert LN. The core principle of the PMOD method involved UV-initiated photodecomposition of spin-coated thin films prepared from a custom photosensitive precursor through a photomask. The exposed regions of the thin films crosslink, and after development with an organic solvent, yielded amorphous patterns. These amorphous patterns are subsequently converted into crystalline LN through a calcination step. Our PMOD approach achieves large-area patterning with feature resolutions down to 30 µm for lines and 70 µm for squares, enabling the fabrication of complex, interconnected structures that were previously unachievable using conventional bottom-up fabrication methods. The patterned thin films consist of phase-pure, polycrystalline, rhombohedral LN, demonstrating that the PMOD method preserves the intrinsic material quality. The patterned LN features exhibit a tunable second harmonic generation (SHG) response, confirming the retention of their nonlinear optical properties. Moreover, the polarization-independent response of polycrystalline LN films opens new opportunities for designing nonlinear optical devices without the need for phase matching or precise polarization control. This PMOD-based approach for scalable and flexible LN patterning could enable the integration of this versatile material system into waveguides, quantum emitters, sensors, frequency converters, and electro-optic modulators.

Several key areas warrant further investigation to optimize the PMOD process and expand its capabilities. For example, systematic studies on precursor chemistry modifications, including alternative photosensitizers and metal-organic complexes, could further enhance the resolution limits of this approach. Similarly, exploring UV light parameters (e.g., power and fluence) and development solvent properties (e.g., polarity and time), in combination with optimized development chemistry, could extend the achievable feature size to sub-10 µm while maintaining ambient processing advantages. Extending PMOD to fabricate other non-centrosymmetric metal oxides (e.g., barium titanate, potassium niobate, and lithium tantalate) could establish a versatile platform for a wide range of nonlinear optical applications. Finally, device-level demonstrations, including integrated waveguides, resonators, and complete photonic circuits, are required to fully validate the commercial potential of PMOD-fabricated LN components and to establish design guidelines for next-generation nonlinear photonic systems.

## Experimental Section

## Materials and Precursor Preparation

All chemicals were used as received without further purification. Lithium niobium ethoxide [$LiNb(OC_2H_5)_6$, 99+%, 5% w/v in ethanol] and 1-phenyl-1,3-butanedione (benzoylacetone, 99%) were purchased from Thermo Fisher Scientific and Sigma-Aldrich, respectively. Anhydrous ethyl alcohol was obtained from Fisher Scientific.

The precursor for photochemical patterning of lithium niobate (LN) was prepared by forming a complex between lithium niobium ethoxide and benzoylacetone. The precursor solution was prepared by mixing 1 mL (0.135 mmol) of lithium niobium ethoxide with either 21.9 mg (0.135 mmol) or 43.8 mg (0.270 mmol) of benzoylacetone in a glass scintillation vial. The vial was wrapped in aluminum foil to prevent exposure to ambient light, and the solution was stirred at 70 °C for 12 h to form the photosensitive precursor.

Silicon wafers and glass slides were cleaned by immersion in acetone under sonication for 2 min, followed by washing with isopropyl alcohol under sonication for an additional 2 min. Substrates were dried and further treated with ozone plasma in a UV/ozone chamber (Bioforce UV/Ozone ProCleaner) for 5 min to remove organic contaminants and improve surface wettability. Thin films of the photosensitive precursors were prepared by spin-coating (500 rpm for 5 s, followed by 2,000 rpm for 15 s) to ensure uniform film thickness and substrate coverage. The coated thin films were baked on a hot plate at 70 °C for 90 s to remove residual solvent while avoiding premature decomposition.

To fabricate LN features, micro-patterned chrome-on-quartz and transparency film photomasks were designed and produced by commercial suppliers and shared research facilities (Artnet Pro and Aggie Fab). The photomasks contained various test patterns, including lines, squares, circles, logos, and complex interconnected shapes with feature sizes ranging from 10 μm to several hundred microns. Thin films of precursors were exposed to UV light (365 nm, 21.1 mW cm$^{-2}$, Cole-Parmer Handheld UV Lamp, 6-Watt, 254/365 nm, UX-09818-07) through the transparent regions of the photomasks for 50 min. The distance between the UV lamp and the substrate was maintained at approximately 5 cm to ensure uniform illumination. All exposures were conducted under ambient atmospheric conditions

Following UV exposure, the patterned substrates were developed by immersion in ethanol for 60 s with gentle agitation, followed by rinsing with fresh ethanol. The developed patterns were calcined in air by heating from room temperature to 650 °C at a rate of 5 °C min$^{-1}$ and holding at 650 °C for 45 min to achieve complete crystallization and removal of organic residues (Cress Electric Furnace, ambient atmosphere).

Optical images of patterned LN thin films were acquired using an Olympus DSX2000 Digital Microscope. The samples were monitored on a motorized stage and focused on acquiring individual images of the smaller areas and stitched images of the larger areas. Optical microscopy images were post-processed using GIMP software (GNU Image Manipulation Program) and Fiji ImageJ to optimize brightness and contrast for improved pattern visualization. These adjustments were applied uniformly across all images and did not affect quantitative dimensional measurements or pattern fidelity assessments.

The patterned substrates were further imaged using scanning electron microscopy (SEM) with a Tescan LYRA-3 focused ion beam (FIB) SEM system. Substrates were mounted on aluminum stubs using carbon tape and sputter-coated with a thin layer of gold-palladium to prevent charging. The composition, crystallinity, and lattice parameters of the grains of the LN thin films were characterized using a Titan Themis 300 S/TEM (scanning/transmission electron microscope) operated at 300 kV. For TEM/STEM analysis, the LN films were carefully detached from the substrates using a plastic spatula and dispersed in ethanol via sonication for 10 min. The resulting suspension was deposited onto Formvar/carbon-coated 300-mesh copper TEM grids (PELCO, Ted Pella Inc., USA) and dried under ambient conditions. Selected area electron diffraction (SAED) data were analyzed using ProcessDiffraction software to generate plots of average pixel intensity versus scattering vector, based on integrated circular diffraction patterns.

Powder X-ray diffraction (XRD) analysis was performed using a D8 Endeavor diffractometer (Bruker AXS). The X-ray source was a 1 kW Cu X-ray tube, operated at 40 kV and 25 mA. The system operated in Bragg-Brentano para-focusing mode, with X-rays diverging through a 1 mm DS slit before striking the sample and converging at a position-sensitive X-ray detector (Lynx-Eye, Bruker-AXS). Data collection was performed using the automated COMMANDER program with a DQL file over a $2\theta$ range of 20-70° with a step size of 0.02° and a counting time of 1 s per step.

Purity, phase, and surface properties were evaluated using Raman spectroscopy. Raman spectra were obtained using a custom-built system comprising a Nikon TE-2000U inverted confocal microscope equipped with a 20× objective (NA = 0.45) and a 785 nm solid-state laser (3 mW). Spectra were collected over the range 100-1000 $cm^{-1}$ with 1 $cm^{-1}$ resolution and 10 s integration time.

UV–visible absorption spectra of precursor solutions and films were measured using a Shimadzu UV-2600 spectrophotometer (Agilent Technologies). Solution measurements were performed in 1 cm pathlength PMMA cuvettes. Crystalline LN thin films on Si were further characterized by spectroscopic ellipsometry using a J.A. Woollam alpha-SE ellipsometer. Ellipsometric parameters ($\Psi$ and $\Delta$) were fitted via a Cauchy model using CompleteEASE software to determine the refractive index.

Second harmonic generation (SHG) activity of the LN patterns was evaluated using an Olympus Fluoview FVMPE-RS multiphoton laser scanning microscope equipped with an Insight X3 laser (tunable range 680-1300 nm, Spectra-Physics) and a Zeiss LSM 880 MP confocal microscope. SHG signals were measured using 5× (NA = 0.15) and 10× (NA = 0.4) objectives, with laser excitation scanned from 840 nm to 1,000 nm as the fundamental wavelengths. The SHG signal was detected using appropriate bandpass filters centered at half the fundamental wavelength. Power dependence measurements were performed by varying the laser power using neutral density filters while monitoring the SHG intensity. Polarization dependence of the SHG response was investigated by rotating a half-wave plate in the excitation beam path to vary the polarization of the incident light. SHG intensity was recorded as a function of polarization angle to assess the isotropy of the polycrystalline films.

## Supporting Information

Supporting Information is available from the Wiley Online Library or from the author.

## Acknowledgements

The authors would like to acknowledge the shared research facilities and individual laboratories for providing access to their equipment and/or assisting with the measurements reported in this manuscript. We sincerely thank Dr. Malea Murphy at the Integrated Microscopy and Imaging Laboratory at the Texas A&M Naresh K. Vashisht College of Medicine (RRID: SCR_021637), Dr. Yordanos Bisrat, Dr. Wilson Seram, and Dr. Sisi Xiang at the Texas A&M University Materials Characterization Core Facility (RRID: SCR_022202), Dr. Jung-Bin Ahn and Dr. Peiran Wei of the Texas A&M University Soft Matter Facility (RRID: SCR_022482), Dr. Joseph Reibenspie at the X-ray Diffraction Facility (Chemistry Department-TAMU, CHE-9807975, CHE-0079822, and CHE-0215838), and Biomedical Engineering shared labs-TAMU (Xin Zhang) for providing access to the equipment and/or training/assisting with the characterization of materials. In addition, the authors would like to thank Professor Marcel Mettlen (UTSW and the Quantitative Light Microscopy Core, a Shared Resource of the Harold C. Simmons Cancer Center, supported in part by an NCI Cancer Center Support Grant, 1P30 CA142543-01.), Dr. Senthil Kumar Boopathi (Professor Karen Wolley's lab-TAMU), Davis Pickett (Professor Dmitry Kurouski's lab-TAMU), Vsevolod Cheburkanov (Professor Vladislav Yakovlev's lab-TAMU), and Atiqul Islam Ahad (Professor Guillermo Aguilar's lab-TAMU) for providing access to equipment and/or assisting with experiments.

## Conflict of Interest

The authors declare no conflict of interest.

## Data Availability Statement

The data that support the findings of this study are available from the corresponding author upon reasonable request.

**Table of Content**

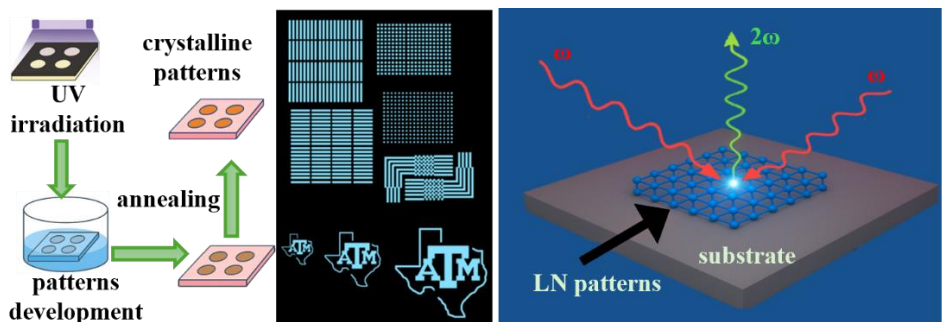

Photochemical metal organic decomposition (PMOD) enables scalable patterning of lithium niobate (LN) metasurfaces under ambient conditions, eliminating the need for cleanrooms or harsh etching. This bottom-up approach produces complex, diverse and size tunable polycrystalline LN patterns. These structures exhibit tunable isotropic second harmonic generation, and require no phase matching, highlighting a simple and scalable route for next-generation photonic device fabrication.

# Supporting Information

## Direct Photochemical Patterning of Lithium Niobate Thin Films for Scalable Nonlinear Optical Metasurfaces

Rana Faryad Ali* and Guillermo Aguilar

J. Mike Walker '66 Department of Mechanical Engineering, Texas A&M University, College Station, TX, 77843 USA

* Email: ranafaryad.ali@tamu.edu

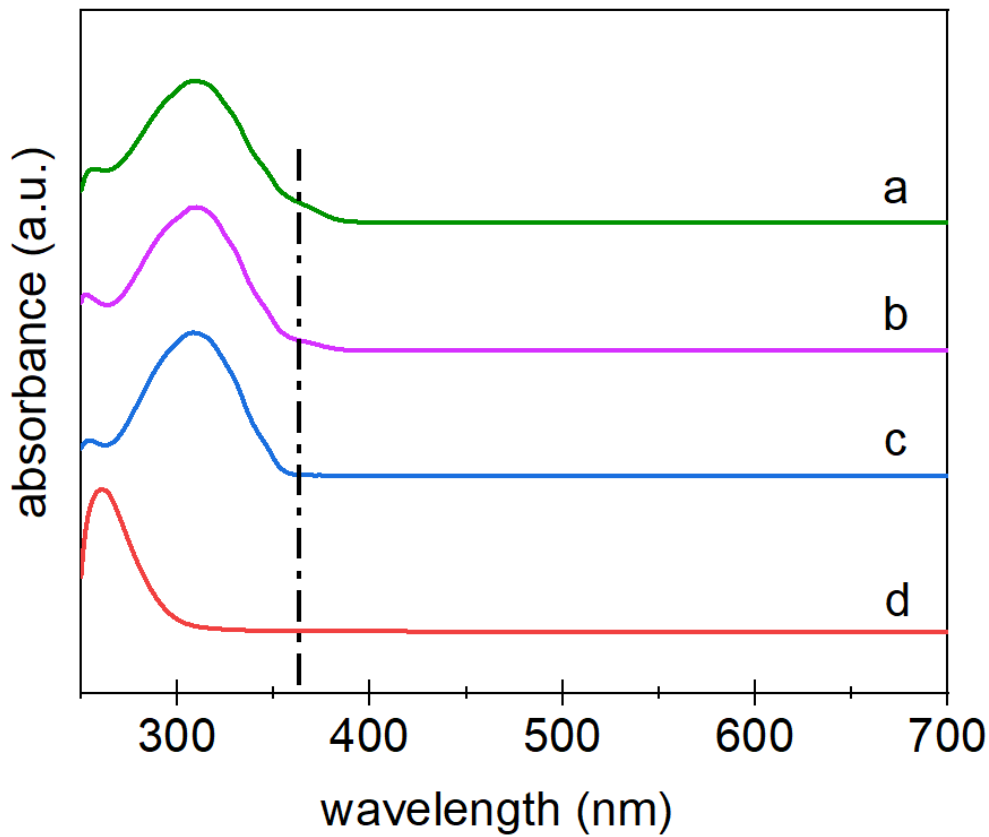

**Figure S1.** UV-visible absorption spectra of the precursor and pristine compounds used for photochemical metal organic deposition (PMOD) fabrication of lithium niobate (LN) thin films. (a) 2:1 molar ratio precursor (lithium niobium ethoxide:benzoylacetone = 2:1) showing a strong absorption band at ~320 nm and an absorption edge at ~365 nm, indicating effective complex formation. (b) 1:1 mole ratio precursor showing a minimal absorption edge at ~365 nm, demonstrating insufficient ligand exchange reaction. (c) UV-visible absorption spectrum of pristine benzoylacetone in ethanol, showing no absorption edge at ~365 nm. (d) Pristine lithium niobium ethoxide exhibiting an absorption band at ~260 nm, characteristic of metal-to-ligand charge transfer transitions in niobium alkoxide complexes.

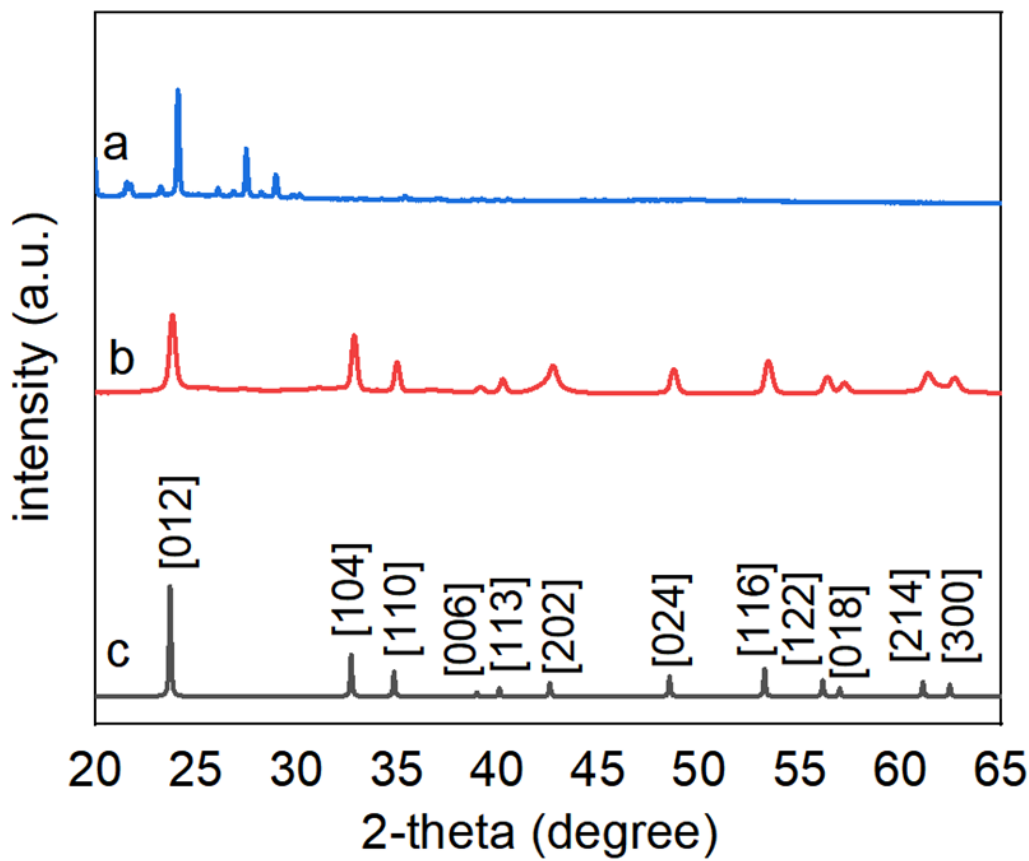

**Figure S2.** Powder X-ray diffraction (XRD) analysis of the lithium niobate (LN) films before and after calcination. (a) XRD patterns of photodecomposed amorphous film before calcination, indicating the absence of crystalline LN. (b) XRD patterns of calcined LN films showing well-defined peaks corresponding to the rhombohedral crystal structure. (c) Standard JCPDS reference patterns for reported LN (JCPDS No. 020-0631).

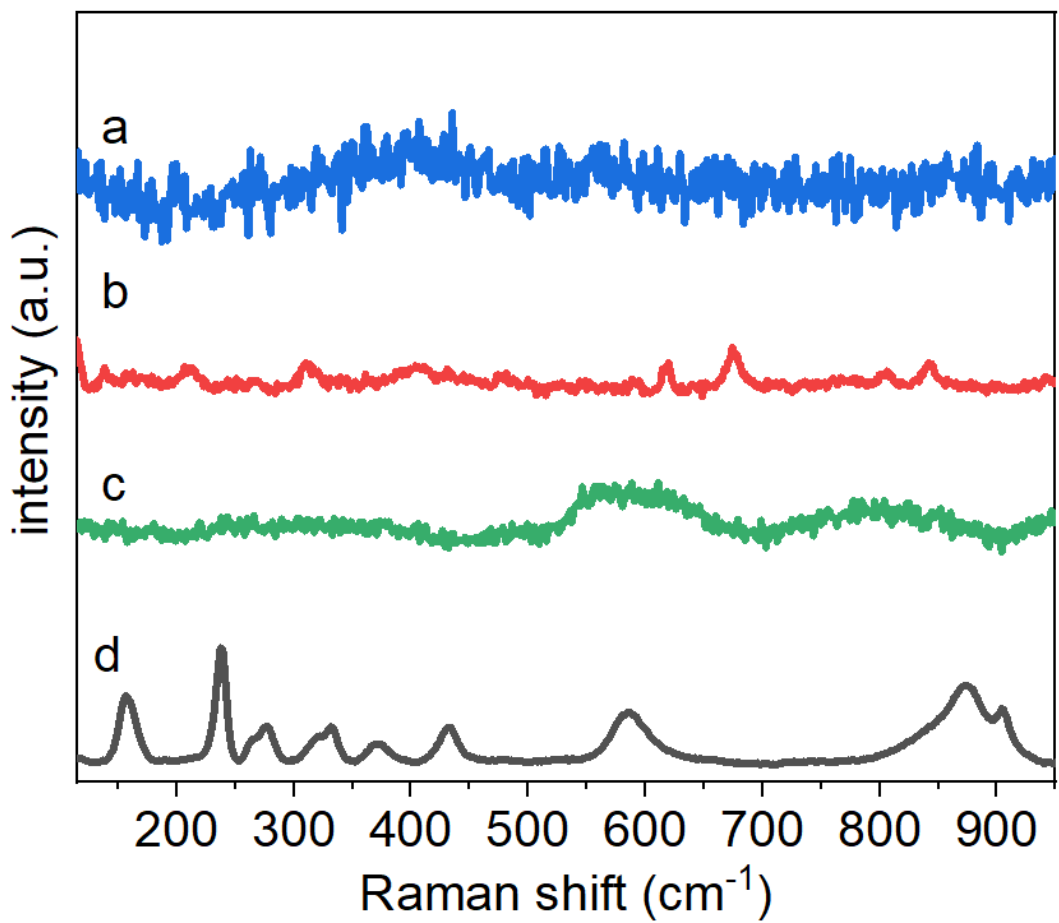

**Figure S3.** Raman spectroscopy analysis of samples during the PMOD fabrication of LN thin film patterns. (a) Pristine lithium niobium ethoxide. (b) 2:1 precursor prepared by mixing benzoylacetone and lithium niobium ethoxide, showing new bands indicative of ligand exchange. (c) Photodecomposed precursor after UV exposure, displaying modified Raman signatures indicative of photodecomposition and the amorphous nature of the LN thin film patterns. (d) Calcined LN Raman spectrum for comparison, illustrating the phase evolution and chemical transformations leading to the formation of crystalline LN.

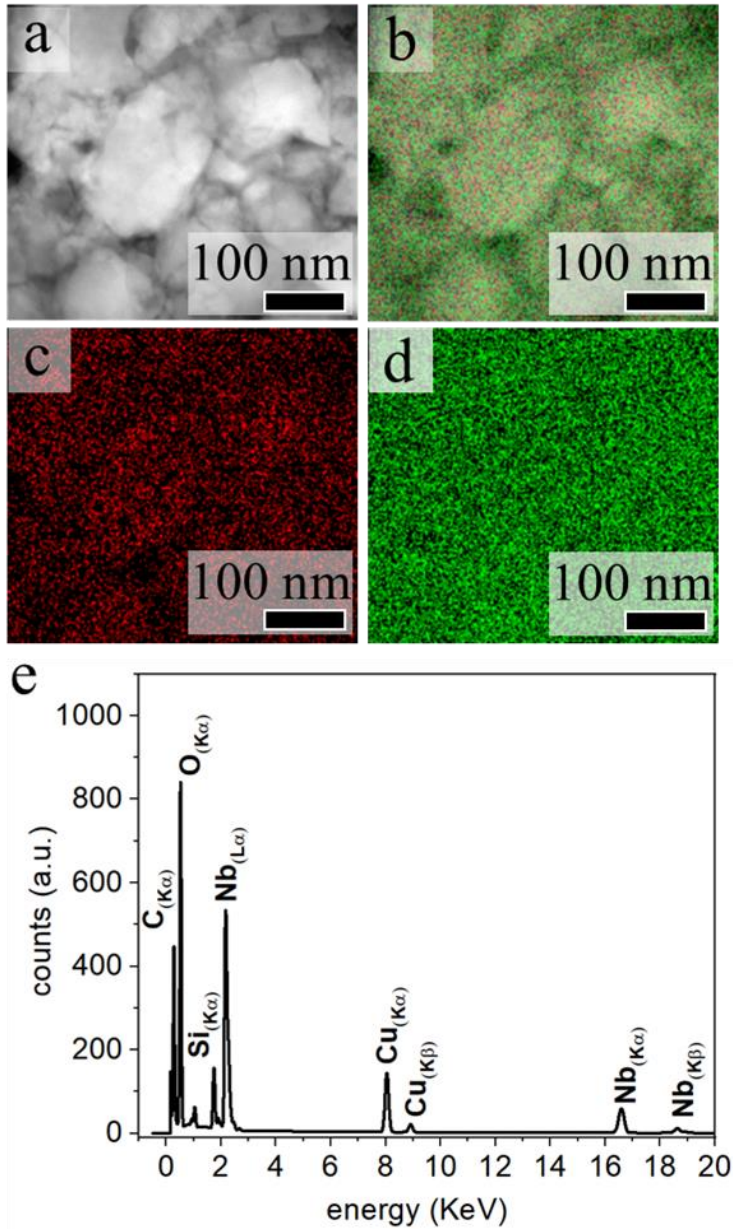

**Figure S4.** Energy-dispersive X-ray spectroscopy (EDS) elemental mapping of particles from the patterned LN thin films. (a) High-angle annular dark-field (HAADF) STEM image showing the morphology of the LN grains. (b) Color-overlapped EDS map displaying the spatial distribution of elements. (c) Niobium elemental map (red). (d) Oxygen elemental map (green), confirming the presence of both elements of LN. (e) A typical EDS spectrum corresponding to the particles, which indicated the presence of Nb and O in the product. The C and Cu contributions were from the TEM grids supporting the sample, and the Si contribution is due to the noise of the EDS detector.

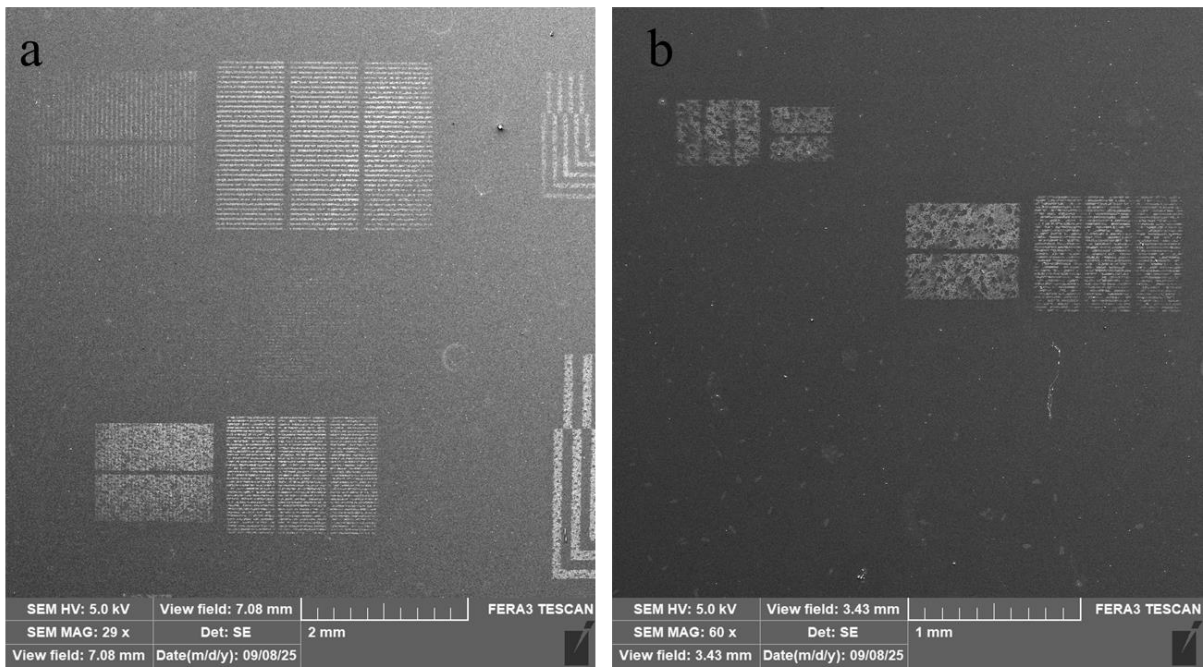

**Figure S5.** Scanning electron microscopy (SEM) analysis of high-resolution line patterns, demonstrating the fabrication limit of the PMOD method. The SEM image shows successful development of line features ≥30 μm in width. Features below 30 μm exhibit poor development, with incomplete pattern transfer and non-uniform edges, illustrating the resolution limitation of the PMOD process under current processing conditions.

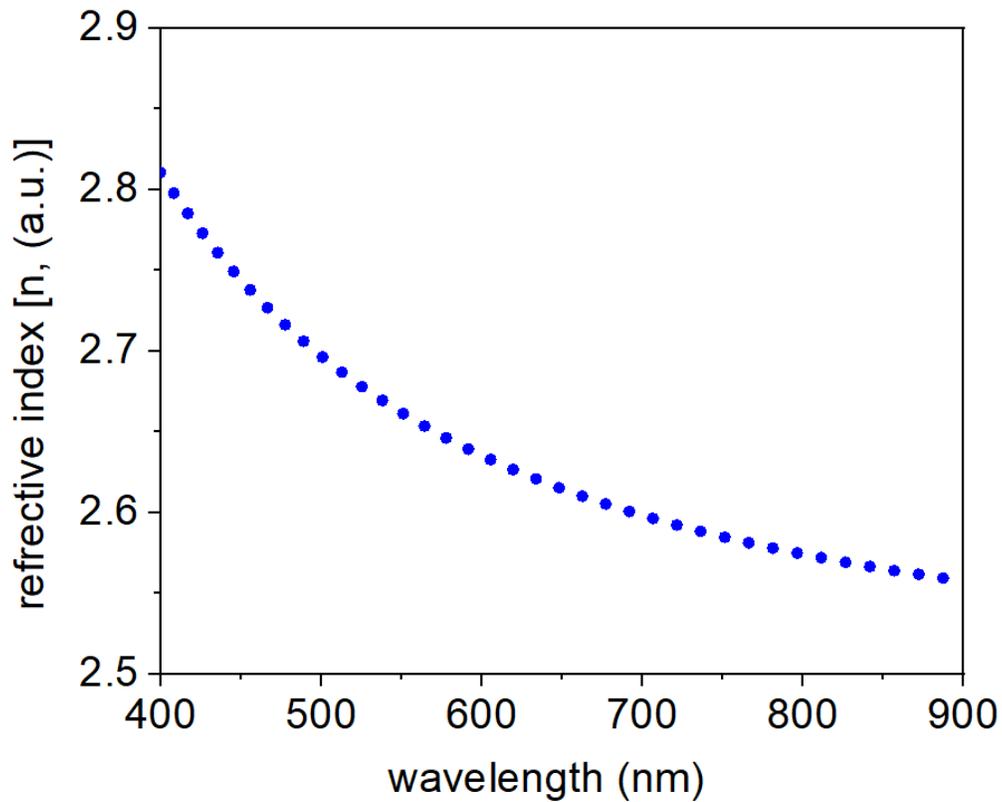

**Figure S6.** Spectroscopic ellipsometry characterization of the optical properties of crystalline LN thin film patterns fabricated on silica. Ellipsometry data showing the refractive index of LN thin films in the visible spectral region (400-900 nm) is wavelength dependent.